\documentclass[a4paper,fleqn,number]{cas-sc}

\usepackage[numbers]{natbib}
\usepackage{subcaption}
\usepackage{indentfirst}
\usepackage{microtype}
\usepackage{siunitx}
\usepackage{placeins}
\usepackage{setspace}
\usepackage{lineno}
\usepackage{xcolor}

\def\tsc#1{\csdef{#1}{\textsc{\lowercase{#1}}\xspace}}
\tsc{WGM}
\tsc{QE}
\tsc{EP}
\tsc{PMS}
\tsc{BEC}
\tsc{DE}

\makeatletter
\renewcommand\subsection{\@startsection
  {subsection}{2}{\z@}%
  {-3.25ex\@plus -1ex \@minus -.2ex}%
  {1.5ex \@plus .2ex}%
  {\normalfont\normalsize\itshape}}
\makeatother

\begin{document}
\modulolinenumbers[1]

\let\WriteBookmarks\relax
\def\floatpagepagefraction{1}
\def\textpagefraction{.001}
\shorttitle{Pitch amplitude and phase effects on turbine power / pitching-energy exchange}
\shortauthors{Z.Zhang et~al.}

\title [mode = title]{Effects of sinusoidal pitch amplitude and phase on power extraction and pitching-energy exchange in a single-blade vertical-axis turbine model}                      

\tnotetext[1]{This project is funded by Iowa Energy Center. YZ acknowledges support from the UC Riverside Winston Chung Global Energy Center and the SoCal OASIS Internal Funding Award.}

\author[1]{Zihan Zhang}[type=editor,
                        auid=000,bioid=1,
                        orcid=0009-0002-2797-9549]
\cormark[1]
\ead{zihan34@iastate.edu}

\credit{Methodology, Software, Validation, Formal analysis, Writing - original draft}

\affiliation[1]{organization={Department of Mechanical Engineering},
                addressline={Iowa State University}, 
                city={Ames},
                postcode={50011}, 
                state={IA},
                country={United States}}

\author[2]{Anchi Li}[style=chinese, orcid=0009-0003-8460-5110]

\ead{ali188@ucr.edu}

\credit{Validation, Formal analysis, Writing - review $\&$ editing}

\author[1]{Qimin Feng}[style=chinese, orcid=0009-0002-1035-7313]
\ead{qmfeng11@iastate.edu}

\credit{Methodology}

\author[2]{Yuanhang Zhu}[style=chinese, orcid=0000-0002-2080-1142]

\ead{yuanhang.zhu@ucr.edu}

\credit{Writing - review $\&$ editing}

\affiliation[2]{organization={Department of Mechanical Engineering},
                addressline={University of California}, 
                city={Riverside},
                postcode={92521}, 
                state={CA},
                country={United States}}

\author[1]{Qiang Zhong}[style=chinese, orcid=0000-0002-8435-5938]
\ead{qzhong1@iastate.edu}
\credit{Conceptualization, Data curation, Writing - review $\&$ editing}

\cortext[cor1]{Corresponding author}

\begin{abstract}
Sinusoidal blade pitching can improve vertical-axis turbine performance, but whether a blade could follow the same motion passively depends on the energy exchanged through pitching. This study experimentally examines 480 prescribed-pitch conditions using a single-blade model following a circular trajectory in a water channel at four tip-speed ratios (TSRs) from 1.25 to 3.14. The amplitude associated with the highest measured power decreases from $16^\circ$ to $12^\circ$ and then $8^\circ$ as TSR increases, with the corresponding phase shifting from $0$ to $\pi/4$. At TSRs of 1.25 and 3.14, the cycle-averaged power coefficient increases from 0.0047 to 0.1037 and from 0.4698 to 0.5948, respectively. Pitching-energy maps show nearly neutral net hydrodynamic pitching work for the highest-performing motions at the two lower TSRs and net energy transfer from the flow to pitching at the two higher TSRs. Virtual damping narrows the positive-energy region, although the higher-TSR best-tested motions retain an energy surplus at nonzero damping. However, net energy transfer from the flow to pitching also occurs for motions with substantial power losses. At the lowest TSR, phase-resolved force measurements and particle image velocimetry link the gain to delayed dynamic-stall-vortex formation and shedding, prolonged favorable loading, and reduced subsequent negative power. At the highest TSR, the flow and loading change locally, with little alteration of the main power-producing event. The joint maps identify performance-enhancing motions with an energetic margin against assumed dissipation, providing candidates for further passive or assisted pitching studies.

\end{abstract}

\begin{graphicalabstract}
\begin{figure*}
    \centering
    \includegraphics[width=1.0\linewidth]{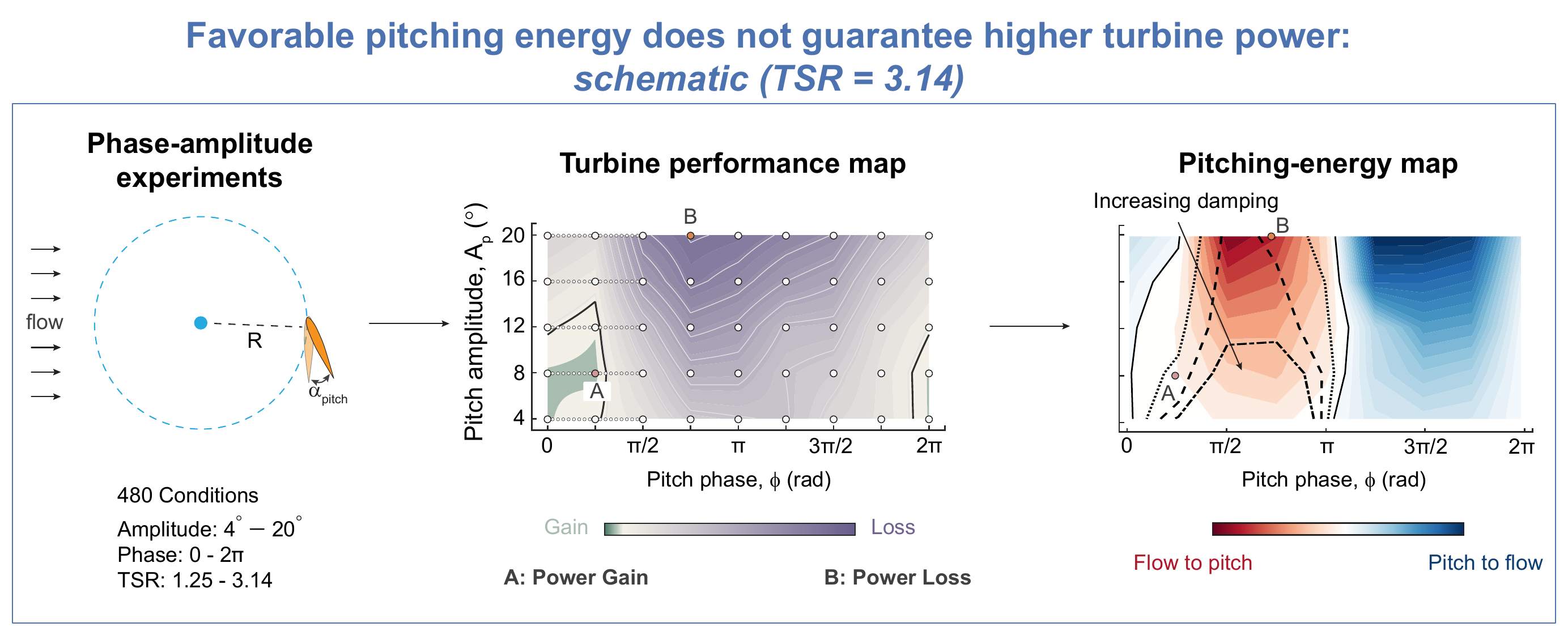}
    \label{fig:graphic abstract}
\end{figure*}

\end{graphicalabstract}

\begin{highlights}
\item Pitch amplitude and phase jointly govern power gains across tip-speed ratios.
\item At low TSR, pitching delays vortex shedding and extends favorable loading.
\item Energy maps screen pitch motions the flow could sustain against damping.
\item The best motions at high TSR receive net energy from the flow through pitching.
\item Net flow-to-pitch energy transfer alone does not ensure a power gain.
\end{highlights}


\maketitle
\doublespacing

\section{Introduction}

Vertical-axis wind turbines (VAWTs) operate without continuous alignment with the incoming wind direction~\cite{xie2017benefits,scherl2020geometric}, but their performance is governed by strongly unsteady blade aerodynamics~\cite{de2022vertical}. During each revolution, the relative flow velocity and effective angle of attack vary with blade azimuth and tip-speed ratio (TSR). These unsteady flow conditions can trigger dynamic stall, during which flow separation and vortex development produce substantial variations in aerodynamic forces and pitching moments~\cite{gardner2023review}. Previous PIV measurements documented the development of dynamic stall~\cite{simao2009visualization,snortland2023cycle}. Complementary experiments characterized its competing effects on power production and load fluctuations~\cite{le2022dynamic,somoano2018dead}, and showed that much of the aerodynamic work occurs during shear-layer growth and vortex formation~\cite{le2023time}. These findings have motivated efforts to regulate the timing and development of separated flow.

Blade pitching offers a direct way to regulate this unsteady response by reshaping the incidence history throughout the turbine cycle. Both intracycle control of rotational velocity~\cite{strom2017intracycle,dave2021simulations} and blade pitching~\cite{elkhoury2015experimental} modify the effective incidence. However, intracycle speed control changes the blade's orbital velocity, whereas pitching allows its orientation to be adjusted while maintaining a prescribed constant orbital angular velocity. Previous studies showed that even a fixed pitch offset can redistribute instantaneous blade loading and alter cycle-averaged performance~\cite{rezaeiha2017effect}. Prescribed cyclic pitching offers further control over the azimuthal distribution of incidence. Erickson et al.~\cite{erickson2011performance} demonstrated improved efficiency and self-starting using mechanically prescribed sinusoidal pitching while examining mean pitch, amplitude, and phase. Benedict et al.~\cite{benedict2016aerodynamics} combined wind-tunnel experiments with computational analysis and showed that turbine performance depends strongly on pitching amplitude and blade kinematics. Jain and Abhishek further established that favorable sinusoidal pitch amplitudes vary with TSR and operating conditions~\cite{jain2016performance}.

The performance benefit of variable pitching depends on how it changes loading over the upstream and downstream portions of the rotation. Sagharichi et al.~\cite{sagharichi2016variable} examined its influence on incidence, separation, and turbine performance, while Chen et al.~\cite{chen2018numerical} related sinusoidal pitching to changes in blade forces, torque, and power fluctuations. Paillard et al.~\cite{paillard2015uranse} investigated harmonic pitch functions using URANS simulations and connected performance changes to the different incidence requirements of the upstream and downstream portions of the rotation. Chen et al.~\cite{chen2019improvement} subsequently modified the downstream portion of a sinusoidal trajectory to reduce resistive torque. More flexible pitch curves have also been optimized by combining computational fluid dynamics with genetic algorithms~\cite{li2018optimization}.


Recent experiments have demonstrated that active pitching can improve net power while limiting load fluctuations. Le Fouest and Mulleners optimized a three-harmonic pitch law using a single-bladed turbine coupled to a genetic algorithm~\cite{le2024optimal}. Their optimization included pitch-actuation power in the net output and considered pitching-moment fluctuations. Their measurements demonstrated changes in force orientation, dynamic-stall-vortex shedding, and downstream energy extraction. Practical variable-pitch mechanisms involve trade-offs between efficiency, mechanism complexity, and force fluctuations~\cite{kirke2011limitations,lazauskas2012modeling}.

These trade-offs have motivated passive variable-pitch concepts, in which the blade motion is driven by the fluid loads rather than by an actuator~\cite{lazauskas2012modeling}. Whether a performance-enhancing motion could be sustained in this way depends on the energy exchanged through the pitching degree of freedom. Energy maps address this question using prescribed-motion experiments: the cycle-averaged work done by the fluid on the body is measured over a grid of kinematic parameters, and regions of net positive work identify motions that the flow can sustain against dissipation. This approach has been used to characterize controlled cylinder oscillations~\cite{morse2009prediction}, prescribed airfoil pitching~\cite{bhat2013stall}, and related nonlinear and aeroelastic responses~\cite{zhu2023flow,menon2020aeroelastic}. In a vertical-axis turbine, the pitching motion is forced at the rotation frequency, so the map is spanned by pitch amplitude and phase rather than amplitude and frequency. Because turbine power depends on the tangential force whereas pitching energy depends on the pitching moment and pitch rate, motions favorable in one need not be favorable in the other. Mapping both over the same amplitudes and phases, with a virtual torsional damper standing in for friction, shows which motions gain power and also receive enough energy from the flow to cover that loss. The maps give the damping at which each motion is energy-balanced; the pitch inertia and stiffness needed to reach its phase are not determined here.

The present study experimentally examines how sinusoidal pitch amplitude and phase jointly affect turbine performance and the pitching-energy balance. A single blade follows a prescribed circular trajectory in a water channel at four TSRs, $\lambda = 1.25$, $1.57$, $2.51$, and $3.14$. A full-cycle coarse phase sweep followed by local refinement yields 480 active-pitch conditions. Performance and pitching-energy contours map net power gains alongside pitching-work exchange, while virtual torsional damping assesses whether favorable combinations retain sufficient net hydrodynamic work to offset the assumed dissipation. Phase-resolved force and power measurements identify where gains and losses arise within the orbit, while PIV measurements at the lowest and highest tested TSRs relate these changes to near-blade vortex development. 

The remainder of this paper is organized as follows. Section 2 describes the experimental facility and single-blade turbine model, prescribed pitching kinematics, force and power calculations, experimental parameter space, pitching-energy analysis, and particle image velocimetry. Section 3 first establishes the fixed-blade baseline and presents the phase–amplitude dependence of turbine performance, followed by an examination of its relationship with pitching-energy exchange. Phase-resolved force and power comparisons then identify the origins of gains and losses within the orbit, and flow-field measurements at the lowest and highest tested TSRs relate these changes to vortex development. Section 4 summarizes the main findings and their implications for assessing sinusoidal pitch schedules.

\section{Methods}

\subsection{Experimental facility and vertical-axis turbine configuration}

\begin{figure}
    \centering
    
    \begin{subfigure}{0.9\textwidth}
        \centering
        \includegraphics[width=\linewidth]{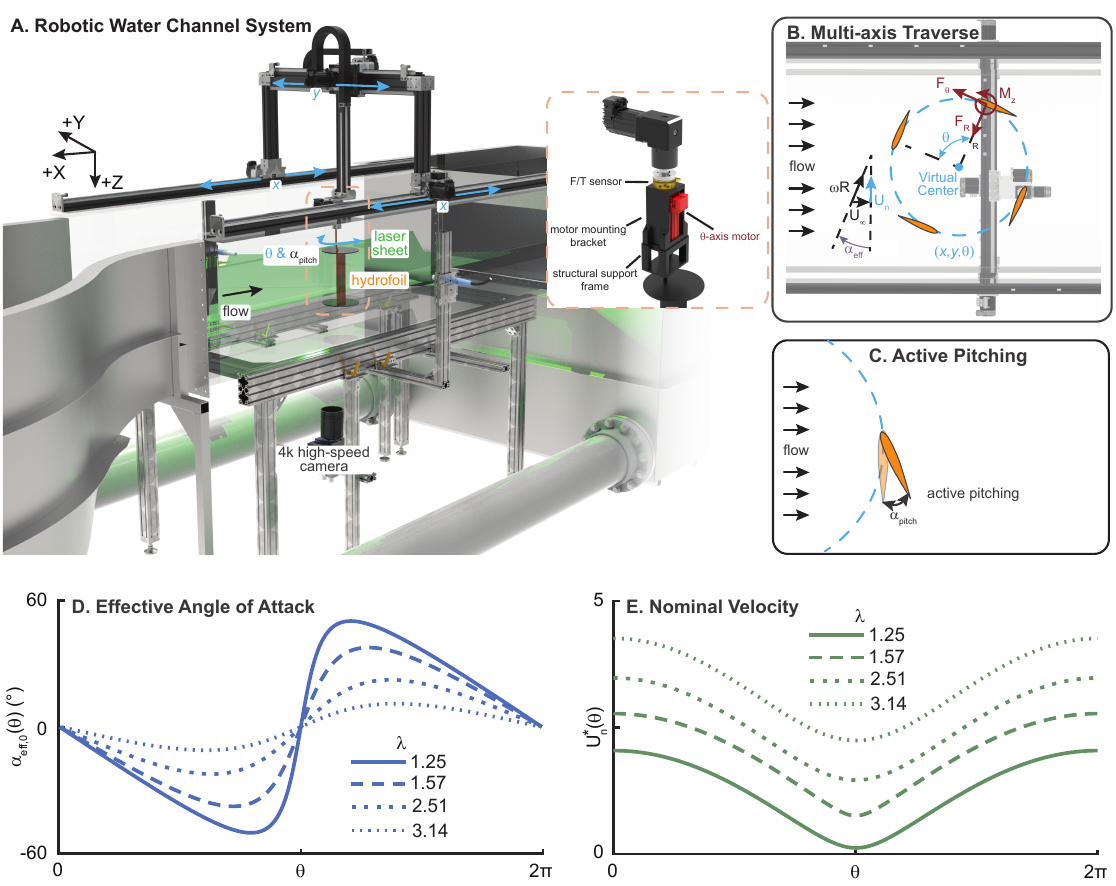}
    \end{subfigure}

    \caption{\textbf{Experimental setup and analysis framework for the active-pitch vertical-axis wind turbine experiments.} \textbf{A.} Robotic water-channel system integrating the multi-axis traverse, hydrofoil, force/torque measurement, and PIV diagnostics. The high-speed camera is used to capture the 2D wake of the dynamic stall at mid-span. \textbf{B.} Multi-axis traverse reproducing the single-blade turbine kinematics about a virtual rotor center. Synchronized $x$- and $y$-axis motions prescribe the circular trajectory of radius $R$, while the rotational axis controls the instantaneous blade orientation. The measured blade loads are resolved into the tangential force, $F_{\theta}$, radial force, $F_\mathrm{R}$, and pitching moment, $M_\mathrm{z}$, following the directions indicated. \textbf{C.} Active-pitch motion imposed through an additional pitch angle, $\alpha_{pitch}$, relative to the baseline blade orientation along the orbital trajectory. \textbf{D.} Effective angle-of-attack profiles, $\alpha_{\mathrm{eff,0}}(\theta)$, over one turbine revolution for the fixed-blade baseline at $\lambda = \omega R/U_{\infty}$ = 1.25, 1.57, 2.51, and 3.14. \textbf{E.} Corresponding normalized effective inflow velocity profiles, $U^{*}_\mathrm{n} = |U_\mathrm{n}| / U_{\infty} = \sqrt{1+2\lambda cos{\theta} +\lambda^2}$, for the same tip-speed ratios.}
    \label{Fig:1}
\end{figure}

As shown in Fig.~\ref{Fig:1}A, experiments were conducted in a closed-loop robotic water channel with a test section measuring $3\,\mathrm{m} \times 1\,\mathrm{m} \times 0.75\,\mathrm{m}$ (length $\times$ width $\times$ height). A single NACA 0018 hydrofoil was mounted beneath the multi-axis traverse system to reproduce the motion of a vertical-axis turbine blade. The blade had a chord length of $c=0.08\,\mathrm{m}$, a span of $s=0.32\,\mathrm{m}$, and an aspect ratio of $AR = s/c = 4.0$. The hydrofoil was 3D-printed from PLA using a Bambu X1 printer and actuated through a $7\,\mathrm{mm}$-diameter carbon-fiber shaft positioned at $p/c=0.2$ from the leading edge. Another carbon-fiber tube with an outer diameter of $10\,\mathrm{mm}$ was installed concentrically around the shaft as a reinforcement sleeve to reduce structural deflection under fluid loading. Circular endplates with a diameter of $d=0.15\,\mathrm{m}$ and a thickness of $t=0.003\,\mathrm{m}$ were attached to the upper and lower ends of the blade to suppress spanwise flow and promote quasi-two-dimensional flow over the blade span.

The orbital trajectory was generated by synchronized motion of three Teknic servomotors (CPM-MCPV-3432P-ELS), with two motors driving the $x$-axis and one driving the $y$-axis. Blade orientation and active pitching were independently controlled by an additional servomotor (CPM-MCPV-2341S-ELS) along the $\theta$-axis. The $\theta$-axis drive assembly was mounted vertically beneath a six-axis force/torque transducer (ATI Mini40 IP65). In this configuration, the force/torque transducer maintained a fixed angular orientation relative to its cable while the blade translated along the prescribed orbital trajectory, thereby avoiding cable winding and enabling continuous load measurements throughout each turbine revolution~\cite{zhang2026reconfigurable}.

Rather than using a mechanically rotating rotor, the turbine kinematics were reproduced by translating a single blade along a prescribed circular trajectory about a virtual turbine center, as illustrated in Fig.~\ref{Fig:1}B. The prescribed turbine diameter was $D=0.40\,\mathrm{m}$, corresponding to an orbital radius of $R=0.20\,\mathrm{m}$ and a chord-to-diameter ratio of $c/D=0.20$. For the present single-blade configuration, the corresponding solidity was $\sigma=Nc/D=0.20$, where $N=1$ is the number of blades~\cite{erickson2011performance,leblanc2018overview}. The relatively large chord-to-diameter ratio was selected to provide pronounced unsteady aerodynamic behavior over the investigated operating conditions, as the chord-to-diameter ratio is known to strongly influence wake development in vertical-axis turbines~\cite{parker2017effect}. The resulting configuration corresponded to a blockage ratio of $17\%$, defined as the ratio of the turbine frontal swept area to the cross-sectional area of the water-channel test section~\cite{le2022dynamic,le2024optimal}.

At this blockage ratio, confinement is expected to increase the measured power coefficients relative to unconfined operation, and values near or above the Betz limit should not be interpreted as unconfined performance. Comparable levels follow from the single-blade experiments of Le Fouest and Mulleners~\cite{le2024optimal} at 12.5\% blockage, where a non-actuated power coefficient of approximately 0.28 at $\lambda = 3.2$ and reported gains by factors of 2.4 to 2.9 imply values between about 0.67 and 0.81. No blockage correction is applied here; following their approach, the analysis focuses on changes relative to the fixed-blade baseline measured under identical conditions.

A laboratory-fixed Cartesian coordinate system $(x,y,z)$ was defined with the positive $x$-direction oriented against the freestream and the positive $z$-direction vertically downward, while the positive $y$-direction was defined to complete a right-handed coordinate system~\cite{zhang2026reconfigurable}. The center of the prescribed circular trajectory was defined as the virtual turbine center, and the instantaneous blade position was described by the azimuthal angle $\theta$, as illustrated in Fig.~\ref{Fig:1}B. Increasing $\theta$ follows the prescribed direction of turbine rotation, and the blade orbital velocity $U_b$ is tangential to the trajectory with a magnitude of $U_b = \omega R$, where $\omega$ is the angular velocity of the virtual rotor. The relative motion between the blade and the incoming flow defines the nominal velocity, $U_{\mathrm{n}}$, and the corresponding effective angle of attack, $\alpha_{\mathrm{eff}}$. The measured hydrodynamic force was decomposed into a tangential component, $F_{\theta}$, acting along the instantaneous direction of blade motion, and a radial component, $F_R$, defined as positive toward the virtual turbine center. The pitching moment about the blade rotational axis was denoted by $M_z$, with its positive direction defined as illustrated in Fig.~\ref{Fig:1}B.

For the fixed-blade baseline, the turbine kinematics were characterized using the tip-speed ratio, $\lambda=\omega R/U_{\infty}$, which represents the ratio between the blade orbital velocity and the freestream velocity. Based on the velocity triangle associated with the prescribed circular motion, the effective angle of attack for the fixed blade can be expressed as:
\begin{equation}
    \alpha_{\mathrm{eff,0}}(\theta)
    =
    -\tan^{-1}
    \left(
    \frac{\sin\theta}
    {\lambda+\cos\theta}
    \right),
    \label{eq:eff aoa_fixed}
\end{equation}
where the sign follows the angle convention defined in Fig.~\ref{Fig:1}B.

The corresponding nominal velocity experienced by the blade, $U_n$, was obtained from the vector combination of the freestream and orbital velocities. Normalized by the freestream velocity, its magnitude is given by:
\begin{equation}
    U_n^*(\theta)
    =
    \frac{|U_n(\theta)|}{U_{\infty}}
    =
    \sqrt{1+2\lambda\cos\theta+\lambda^2}.
\end{equation}

The resulting azimuthal variations of $\alpha_{\mathrm{eff}}$ and $U_n^*$ for the investigated tip-speed ratios are shown in Fig.~\ref{Fig:1}D and E, respectively. Increasing $\lambda$ reduces the range of effective angle of attack while increasing the relative contribution of the blade orbital velocity to the nominal inflow~\cite{strom2017intracycle}.

For all experiments, the free-stream velocity was maintained at $U_{\infty}=0.20\,\mathrm{m/s}$ and independently verified using an ultrasonic flow meter (TUF-2000B). This condition corresponds to a chord-based Reynolds number of $Re = \rho U_{\infty} c / \mu \approx 1.5 \times 10^4$. Following the same experimental setup~\cite{zhang2026inertial}, the ultrasonic flow sensor was installed approximately $0.5\,\mathrm{m}$ downstream of the contraction exit, while the foil was located approximately $1.0\,\mathrm{m}$ further downstream. The prescribed freestream velocity was verified with the hydrofoil stationary and continuously monitored during rotation. The measurement yielded a time-averaged inflow velocity of $U_{\infty}=0.200\pm0.007\,\mathrm{m/s}$, confirming a stable upstream flow condition and minimizing concern that the downstream orbital motion affected the velocity measurement. Forces and moments were sampled at $4000\,\mathrm{Hz}$ using the six-axis force/torque transducer and a 16-bit data-acquisition system (NI PCIe-6363). Motor commands and force/torque acquisition were synchronized through a Simulink-based real-time control program, enabling phase-resolved measurements over repeated turbine revolutions.

\subsection{Blade kinematics and active-pitching strategies}

The blade trajectory and active-pitch motion were prescribed through synchronized commands of the multi-axis traverse. As illustrated in Fig.~\ref{Fig:1}B, the azimuthal position of the blade is described by $\theta$, with the corresponding circular trajectory generated through the synchronized $x$- and $y$-axis motions. For a constant angular velocity, the azimuthal position evolves as $\theta(t) = \omega t$. Then, the translational commands are given by:
\begin{equation}
    x(t) = R\sin\theta(t), \qquad
    y(t) = R\cos\theta(t),
\end{equation}
with the orbital radius of $R = 0.20\,\mathrm{m}$. For the fixed-blade baseline, no additional pitching motion is imposed relative to the nominal blade orientation along the circular trajectory, such that $\alpha_p(\theta)=0$. With active pitching, an additional angular displacement is imposed as:
\begin{equation}
    \alpha_p(\theta)
    =
    A_p\sin(\theta+\phi),
\end{equation}
where $A_p$ is the single-sided pitch amplitude and $\phi$ is the phase offset between the active-pitch motion and the blade azimuthal position.

The imposed pitch motion directly modifies the instantaneous blade incidence relative to the nominal inflow. Accordingly, the effective angle of attack under active pitching is obtained by correcting the fixed-blade value by the prescribed pitch angle, $\alpha_{\mathrm{eff}}(\theta) = \alpha_{\mathrm{eff},0}(\theta) + \alpha_p(\theta)$. Substituting the prescribed pitching law gives:
\begin{equation}
    \alpha_{\mathrm{eff}}(\theta)
    =
    -\tan^{-1}
    \left(
    \frac{\sin\theta}
    {\lambda+\cos\theta}
    \right)
    +
    A_p\sin(\theta+\phi).
    \label{eq:eff aoa}
\end{equation}

\subsection{Force decomposition and power calculation}
The hydrodynamic loads acting on the blade were measured using the six-axis force/torque transducer described in Section~2.1. The measured in-plane forces were resolved into the tangential and radial components defined in Fig.~\ref{Fig:1}B. Tangential force is defined as positive in the direction of blade motion and radial force as positive toward the virtual turbine center. Accordingly, the force transformation from the laboratory-fixed coordinate system to the turbine coordinate system is given by~\cite{snortland2025experimental}:
\begin{equation}
    F_{\theta}(\theta) = F_x(\theta)\cos\theta - F_y(\theta)\sin\theta, \qquad
    F_R(\theta) = -F_x(\theta)\sin\theta - F_y(\theta)\cos\theta,
\end{equation}
where $F_x$ and $F_y$ are the measured force components in the laboratory-fixed coordinate system. The pitching moment about the blade rotational axis is denoted by $M_z$, with the positive direction defined in Fig.~\ref{Fig:1}B.

To facilitate comparison of the blade loading across different operating conditions, the tangential and radial forces were non-dimensionalized using the blade orbital velocity ($U_b$), together with the blade chord ($c$) and span ($s$). The corresponding force coefficients are defined as:
\begin{equation}
    C_{F, \theta}(\theta)
    =
    \frac{F_{\theta}(\theta)}
    {0.5\rho U_b^2 sc},
    \qquad
    C_{F, R}(\theta)
    =
    \frac{F_R(\theta)}
    {0.5\rho U_b^2 sc},
\end{equation}

The instantaneous mechanical power associated with the orbital motion is determined from the tangential force and the blade orbital velocity, $P_{\mathrm{orb}}(\theta) = F_{\theta}(\theta)R\omega$. A positive value of $P_{\mathrm{orb}}$ indicates that the hydrodynamic force acts in the direction of blade motion and therefore contributes positively to turbine power generation.

For the active-pitch cases, part of the extracted power is required to actuate the prescribed pitching motion. The instantaneous pitch-actuation power is evaluated as $P_{\mathrm{pitch}}(\theta) = M_z(\theta)\dot{\alpha}_p(\theta)$, where the angular velocity associated with the prescribed pitch law in Section~2.2 is $\dot{\alpha}_p(\theta) = A_p\omega\cos(\theta+\phi)$. The instantaneous net power of the actively pitched blade is therefore defined as:
\begin{equation}
    P_{\mathrm{net}}(\theta) = F_{\theta}(\theta)R\omega + M_z(\theta)\dot{\alpha}_p(\theta).
\end{equation}

The two terms represent power generated through orbital motion and hydrodynamic power associated with active pitching, respectively. For the fixed-blade baseline, $\dot{\alpha}_p=0$, and the net power reduces directly to the orbital contribution.

To compare turbine performance across operating conditions, the instantaneous net power was normalized by the kinetic energy flux through the turbine swept area~\cite{le2024optimal}:
\begin{equation}
    C_P(\theta) = \frac{P_{\mathrm{net}}(\theta)}{0.5\rho U_{\infty}^{3}A_{\mathrm{swept}}},
    \label{eq:net power}
\end{equation}
where $A_{\mathrm{swept}}$ is the product of turbine diameter and foil's span, $A_{\mathrm{swept}} = Ds$. The cycle-averaged power coefficient was subsequently calculated over $n$ complete turbine revolutions as:
\begin{equation}
    \bar{C}_P = \frac{1}{2\pi n}\int_{0}^{2\pi n}C_P(\theta)\,d\theta,
\end{equation}
which is used as the primary performance metric for comparing the fixed-blade and active-pitch configurations throughout this study.




\subsection{Experimental parameter space}

\begin{table}[width=1.0\textwidth,cols=5,pos=h]
\caption{Experimental parameter space for the active-pitch investigation. The refined phase sweep was centered on the optimal phase identified from the initial sweep, with phase values wrapped periodically over 
$0\leq\phi<2\pi$.}
\label{tbl1}
\begin{tabular*}{\tblwidth}{@{} CCCCC @{} }
\toprule
$\lambda\;(=\omega R/U_{\infty})$
& $A_p$
& $\phi_{\mathrm{initial}}$
& $\phi_{\mathrm{coarse,best-tested}}$
& $\phi_{\mathrm{refined}}$ \\
\midrule

1.25
& \multirow{4}{*}{$[4^\circ,\;8^\circ,\;12^\circ,\;16^\circ,\;20^\circ]$}
& \multirow{4}{*}{
\shortstack{
$[0,\;\pi/4,\;\pi/2,\;3\pi/4,$\\
$\pi,\;5\pi/4,\;3\pi/2,\;7\pi/4]$
}}
& 0
& \multirow{4}{*}{
$\mathrm{mod}\!\left(
\phi_{\mathrm{coarse,best-tested}}
\pm k\pi/36,\;2\pi
\right),\;k=1,\ldots,8$
} \\

1.57
&
&
& 0
&  \\

2.51
&
&
& $\pi/4$
&  \\

3.14
&
&
& $\pi/4$
&  \\

\bottomrule
\end{tabular*}
\end{table}

A two-stage phase--amplitude search strategy was employed to identify the active-pitch kinematics that maximize turbine performance at each tip-speed ratio. Four tip-speed ratios were investigated, $\lambda=1.25$, 1.57, 2.51, and 3.14, based on four orbital frequencies $f=0.20$, 0.25, 0.40, and 0.50 Hz. For each $\lambda$, an initial coarse sweep was performed using eight uniformly distributed pitch phases over one turbine revolution, $\phi_\mathrm{initial} = 0,\,\pi/4,\,\pi/2,\,3\pi/4,\,\pi,\,5\pi/4,\,3\pi/2,\,7\pi/4$, combined with five prescribed pitch amplitudes, $A_p = 4^\circ,\,8^\circ,\,12^\circ,\,16^\circ,\,20^\circ$. This initial parameter sweep therefore consisted of 40 active-pitch cases at each tip-speed ratio.

The phase--amplitude combination producing the highest cycle-averaged power coefficient from the coarse sweep was then used to define a refined phase search. Sixteen additional phase values were introduced in the neighborhood of the coarse best-tested case, while the same five pitch amplitudes were retained. As an example, for $\lambda=1.25$, the coarse best-tested case occurred near $\phi=0$. The refined search therefore covered the interval surrounding $\phi=0$ using:
\begin{equation*}
\begin{split}
    \phi_\mathrm{refined} ={}&
    \frac{\pi}{36},\,
    \frac{\pi}{18},\,
    \frac{\pi}{12},\,
    \frac{\pi}{9},\,
    \frac{5\pi}{36},\,
    \frac{\pi}{6},\,
    \frac{7\pi}{36},\,
    \frac{2\pi}{9}, \\
    &
    \frac{16\pi}{9},\,
    \frac{65\pi}{36},\,
    \frac{11\pi}{6},\,
    \frac{67\pi}{36},\,
    \frac{17\pi}{9},\,
    \frac{23\pi}{12},\,
    \frac{35\pi}{18},\,
    \frac{71\pi}{36}.
\end{split}
\end{equation*}

These values correspond to $5^\circ$ increments extending from $320^\circ$ to $40^\circ$ across the periodic boundary at $\phi=0$. The same refinement procedure was applied independently at each $\lambda$, with the refined phase window centered around the corresponding optimum identified from the coarse sweep.

Combining the eight coarse phases and 16 locally refined phases resulted in 24 pitch phases at each tip-speed ratio. Together with the five pitch amplitudes, a total of 120 active-pitch cases were evaluated for each $\lambda$, yielding 480 active-pitch cases across the four investigated operating conditions. A fixed-blade case, corresponding to $A_p=0$, was additionally tested at each tip-speed ratio and used as the baseline for performance comparisons. The full experimental parameter space is summarized in Table~\ref{tbl1}.

Each operating condition was acquired in a single experimental trial consisting of 20 consecutive turbine revolutions. To minimize transient effects associated with the initiation and termination of the prescribed motion, only cycles 3--18 were retained for subsequent analysis. The resulting 16 cycles were phase averaged to obtain the azimuthally resolved forces, moments, and power quantities reported throughout this study.

\subsection{Force-based energy-map analysis}

\label{sec:energy_map}

To evaluate the power exchanged through the pitching motion, a force-based energy-map analysis was performed following the approach of Zhu et al.~\cite{zhu2020nonlinear}. Whereas turbine performance is evaluated using the swept area $Ds$ as the reference area, the pitching-energy analysis below adopts the blade planform area $cs$ to characterize energy exchange in the pitching degree of freedom. The instantaneous pitching power defined in Section~2.3, $P_{\mathrm{pitch}}(\theta) = M_z(\theta)\,\dot{\alpha}_p(\theta)$, represents the rate of work done by the hydrodynamic moment on the pitching blade. Positive $P_{\mathrm{pitch}}$ denotes energy transfer from the flow to the pitching motion; negative values denote the reverse transfer.

To assess the pitching-energy balance under prescribed dissipative losses, a virtual torsional damper with a damping coefficient $b$ was introduced along the pitch axis. The instantaneous power dissipated by the damper is evaluated as:
\begin{equation}
    P_{\mathrm{damp}}(\theta) = b\,\dot{\alpha}_p^{2}(\theta).
    \label{eq:Pdamp}
\end{equation}
Accordingly, the instantaneous net power associated with the pitching motion is defined as:
\begin{equation}
    P_{\mathrm{net,pitch}}(\theta) = P_{\mathrm{pitch}}(\theta) - P_{\mathrm{damp}}(\theta).
    \label{eq:Pnetpitch}
\end{equation}

Positive and negative values of $P_{\mathrm{net,pitch}}$ indicate an instantaneous surplus and deficit, respectively, of hydrodynamic pitching power relative to the assumed damping loss. Its cycle average quantifies whether net hydrodynamic pitching work is sufficient to offset that loss.

To facilitate comparison across operating conditions, the damping coefficient and the net pitching power were non-dimensionalized using the blade chord ($c$) and span ($s$), following Zhu et al.~\cite{zhu2020nonlinear}. The corresponding non-dimensional damping coefficient and net pitching power coefficient are defined as:
\begin{equation}
    b^{*} = \frac{b}{0.5\rho U_\infty c^{3} s}, \qquad
    C_{P,\mathrm{net,pitch}}(\theta) = \frac{P_{\mathrm{net,pitch}}(\theta)}{0.5\rho U_\infty^{3} c\, s}.
    \label{eq:bstar_Cpnetpitch}
\end{equation}
The cycle-averaged net pitching power coefficient was subsequently calculated over the $n = 16$ retained turbine revolutions (Section~2.4) as:
\begin{equation}
    \bar{C}_{P,\mathrm{net,pitch}} = \frac{1}{2\pi n}\int_{0}^{2\pi n} C_{P,\mathrm{net,pitch}}(\theta)\, d\theta.
    \label{eq:Cpnetpitch_bar}
\end{equation}


For each tip-speed ratio, $\bar{C}_{P,\mathrm{net,pitch}}$ was evaluated for all 120 phase--amplitude combinations listed in Table~1. Four damping coefficients were examined, $b = 0$, $0.03$, $0.06$, and $0.09~\mathrm{N\,m\,s/rad}$, corresponding to $b^{*} = 0$, $1.83$, $3.66$, and $5.49$, respectively. For each damping coefficient, the $\bar{C}_{P,\mathrm{net,pitch}} = 0$ contour marks equality between cycle-averaged hydrodynamic pitching power and the assumed damping loss, separating positive and negative energy balances.

Interpreting these maps in terms of passive pitching rests on three assumptions. First, because the pitching moment is forced by the orbital incidence variation, the steady response of a passively pitching blade is taken to be periodic at the rotation frequency; self-excited responses at other frequencies are not considered. Second, the response is approximated by its fundamental harmonic. This is appropriate when the pitch inertia is large enough that the response to the higher harmonics of the moment is inertia-dominated; a stiffness- or damping-dominated blade would instead follow the forcing waveform more closely. Third, mechanical losses are represented by a linear viscous damper. 
Under these assumptions, the cycle-averaged balance in Eq.~(\ref{eq:Cpnetpitch_bar}) is a necessary but not sufficient condition for passive operation.

Consider a passive pitch axis with inertia $I$, stiffness $\kappa$, and damping $b$, driven by the hydrodynamic moment~\cite{zhu2021nonlinear}: $I\ddot{\alpha}_p + b\,\dot{\alpha}_p + \kappa\,\alpha_p = M_z$. A steady sinusoidal response at the rotation frequency satisfies two balances. The component of $M_z$ in phase with the pitch rate balances the damping, which is the energy balance of Eq.~(\ref{eq:Cpnetpitch_bar}). The component in phase with the pitch angle balances the net stiffness, $\kappa - I\omega^{2}$, and sets the phase. A passive blade with a given damping can therefore respond only at points on the corresponding $\bar{C}_{P,\mathrm{net,pitch}} = 0$ contour, with its inertia and stiffness selecting the position along that contour. Because this selection depends on the ratio of the pitch natural frequency to the rotation frequency, it is specific to the operating condition. The present analysis evaluates the first balance; the second requires the in-phase moment component and is left for future work.

\subsection{Particle image velocimetry}



Two-dimensional particle image velocimetry (PIV) was used to characterize the phase-resolved vortex dynamics associated with the fixed-blade and best-tested active-pitch turbine configurations. The flow was seeded with $60\,\mu\mathrm{m}$ polyamide particles and illuminated by a $10\,\mathrm{W}$ laser sheet (LWPIV-XH-10) positioned at the blade mid-span. Because the field of view required to resolve the blade trajectory and its surrounding wake exceeded that of a single camera, two high-speed cameras (Photron Nova R3-4K) were arranged in a side-by-side configuration. Images were acquired at a resolution of $4096\times2304$ pixels and synchronized with the prescribed turbine motion.

The two camera views were calibrated and combined in LaVision DaVis 10 using the side-by-side calibration and image/vector-field stitching procedure. Both cameras were mapped to a common physical coordinate system, after which the geometrically corrected fields were stitched to form a single continuous field of view encompassing the region of interest along the turbine trajectory. This procedure enabled the near-blade vortex structures and their downstream evolution to be tracked over a substantially larger spatial extent than could be captured by either camera individually.

PIV measurements were performed at the lowest and highest investigated tip-speed ratios, $\lambda=1.25$ and $3.14$, to examine the flow mechanisms associated with the different active-pitch responses identified from the performance measurements. As summarized in Table~\ref{tbl2}, the fixed-blade condition was compared with the corresponding best-tested active-pitch condition at each tip-speed ratio. The optimal pitching parameters were $(\phi,A_p)=(0,16^\circ)$ at $\lambda=1.25$ and $(\pi/4,8^\circ)$ at $\lambda=3.14$.

Owing to the relatively low rotational frequency of the turbine and the high image resolution, $1000$ consecutive images were acquired over a single complete turbine revolution for each condition. From each sequence, $100$ phase-resolved fields distributed over the turbine cycle were retained for subsequent flow-field analysis. Eight representative azimuthal positions were then selected to illustrate the evolution of the near-blade vortex structures and their relationship with the measured tangential loading and instantaneous power in Figs.~\ref{FIG:6} and \ref{FIG:7}. The resulting velocity fields were processed in LaVision DaVis 10 by using a standard cross-correlation algorithm with overlapping $128 \times 128$ pixel interrogation windows, and the corresponding spanwise vorticity, $\omega_z$, was used to identify and track the formation, attachment, and shedding of the dominant vortex structures over the turbine cycle.

\begin{table}[width=.9\textwidth,cols=4,pos=h]
\caption{Turbine conditions selected for PIV measurements.}
\label{tbl2}
\begin{tabular*}{\tblwidth}{@{} CCC @{} }
\toprule
$\lambda\;(=\omega R/U_{\infty})$ & Fixed-blade & Best-tested active pitch([$\phi$, $A_p$])\\
\midrule

1.25 & zero-pitch & [0, $16^\circ$] \\
3.14 & zero-pitch & [$\pi/4$, $8^\circ$] \\

\bottomrule
\end{tabular*}
\end{table}

\section{Results}

\subsection{Fixed-blade performance across TSRs}

\begin{figure}
	\centering
        \begin{subfigure}{1.0\textwidth}
            \centering
	       \includegraphics[width=0.75\linewidth]{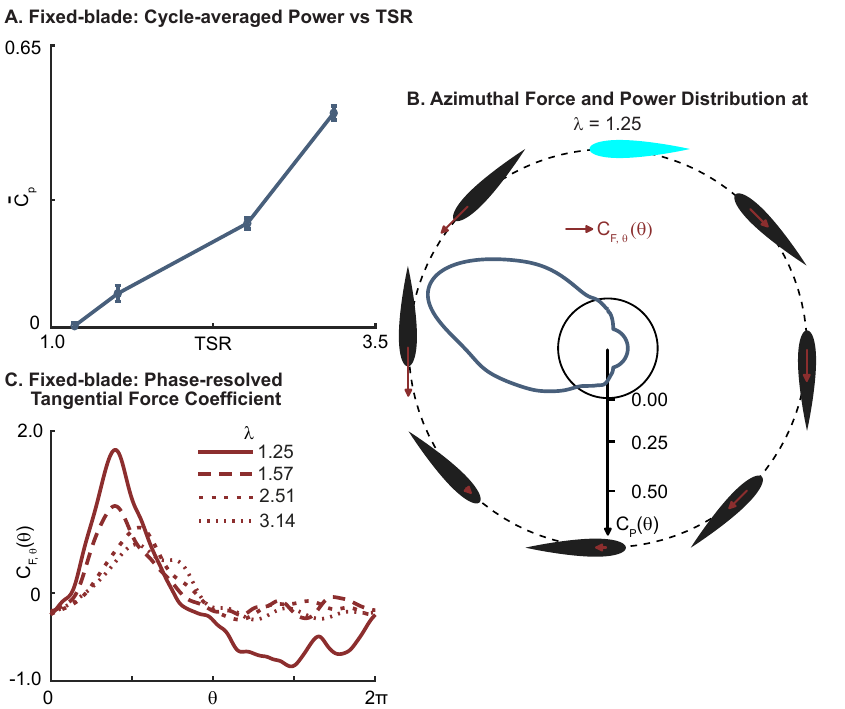}
        \end{subfigure}
    
	\caption{\textbf{Fixed-blade performance and loading characteristics over the investigated tip-speed-ratio range.} \textbf{A.} Cycle-averaged power coefficient, $\bar{C}_P$, of the fixed-blade turbine at $\lambda$ = 1.25, 1.57, 2.51, and 3.14. Each marker represents the mean of the cycle-averaged power coefficients obtained from 16 turbine revolutions, and the error bars indicate one standard deviation across these cycles. \textbf{B.} Azimuthal tangential-force and power distributions for the representative low-tip-speed-ratio condition, $\lambda$ = 1.25. The red arrows indicate the direction and relative magnitude of the tangential force coefficient, $C_{F, \theta}(\theta)$, at representative blade positions, while the polar profile represents the corresponding instantaneous power coefficient, $C_{P}(\theta)$, throughout the turbine revolution. The cyan foil identifies the reference position, $\theta = 0$. \textbf{C.} Phase-resolved tangential force coefficient, $C_{F, \theta}(\theta)$, for the four fixed-blade operating conditions investigated in the present study. \protect}
	\label{FIG:2}
\end{figure}

The fixed-blade measurements first establish the baseline turbine performance over the investigated tip-speed-ratio range. As shown in Fig.~\ref{FIG:2}A, the cycle-averaged power coefficient, $\bar{C}_P$, increases monotonically with $\lambda$ over the tested conditions, with the lowest power extraction occurring at $\lambda=1.25$ and the highest at $\lambda=3.14$. The error bars, obtained from the cycle-to-cycle variation of the 16 retained turbine revolutions, remain relatively small compared with the variation in $\bar{C}_P$ across tip-speed ratios, indicating repeatable cycle-averaged performance. The increase in $\bar{C}_P$ is consistent with the corresponding changes in the fixed-blade kinematics shown in Fig.~\ref{Fig:1}D,E, where increasing $\lambda$ reduces the excursion of the effective angle of attack while increasing the nominal velocity experienced by the blade~\cite{li2018optimization}.

The phase-resolved tangential loading provides further insight into the origin of this variation in cycle-averaged performance. Figure~\ref{FIG:2}C shows that the fixed blade experiences strongly non-uniform tangential loading over each turbine revolution. At $\lambda=1.25$, $C_{F, \theta}$ exhibits a pronounced positive peak during the first portion of the cycle, followed by an extended region of negative tangential loading over the later portion of the orbit. Because positive $F_\theta$ acts in the direction of blade motion, these two regions contribute oppositely to the instantaneous turbine power. As $\lambda$ increases, the magnitude of the positive peak decreases and the phase-resolved loading becomes progressively less oscillatory, while the negative-loading excursion is also substantially reduced.

A similar dependence of tangential loading on tip-speed ratio was reported by Le Fouest and Mulleners~\cite{le2022dynamic}, who observed that decreasing $\lambda$ produced a larger and earlier peak in the tangential force coefficient, followed by a pronounced negative excursion at sufficiently low tip-speed ratios. Their time-resolved flow measurements attributed this behavior to increasingly severe dynamic stall: at lower $\lambda$, the larger variation in $\alpha_\mathrm{eff}$ (Fig.~\ref{Fig:1}D) promotes the earlier formation and growth of a stronger dynamic-stall vortex, which temporarily enhances tangential loading before vortex shedding and subsequent flow separation lead to a rapid load reduction. At higher tip-speed ratios, the dynamic-stall vortex becomes weaker and less coherent, resulting in reduced load fluctuations and a more uniform force response over the turbine cycle~\cite{le2022dynamic}. The similar evolution observed in Fig.~\ref{FIG:2}C suggests that dynamic-stall-related flow development may also play an important role in the present fixed-blade loading response.

The spatial distribution of the blade loading is illustrated more directly in Fig.~\ref{FIG:2}B for the representative low-tip-speed-ratio case, $\lambda=1.25$. The tangential force coefficient varies substantially in both magnitude and direction as the blade progresses around the virtual rotor, producing a strongly nonuniform power response over the turbine cycle. The corresponding azimuthal power distribution further demonstrates that power production is concentrated within a limited portion of the turbine orbit, whereas unfavorable tangential loading over other azimuthal positions leads to reduced and eventually negative instantaneous power. In particular, the large variation in $C_{F, \theta}$ results in a strongly localized positive-power region followed by a substantial reduction, and eventual reversal, of the instantaneous power contribution later in the cycle.

The fixed-blade results therefore show that turbine performance is governed not only by the magnitude of the hydrodynamic loading, but also by its distribution over the azimuthal cycle. In particular, the coexistence of favorable and unfavorable tangential loading suggests that modifying the blade incidence at selected portions of the orbit may provide a means of redistributing the power-producing regions. This motivates the use of active pitching, for which both the timing and magnitude of the prescribed pitch motion can be varied systematically.

\subsection{Phase–amplitude dependence of turbine performance}

\begin{figure}
	\centering
        \begin{subfigure}{1.0\textwidth}
            \centering
	       \includegraphics[width=0.9\linewidth]{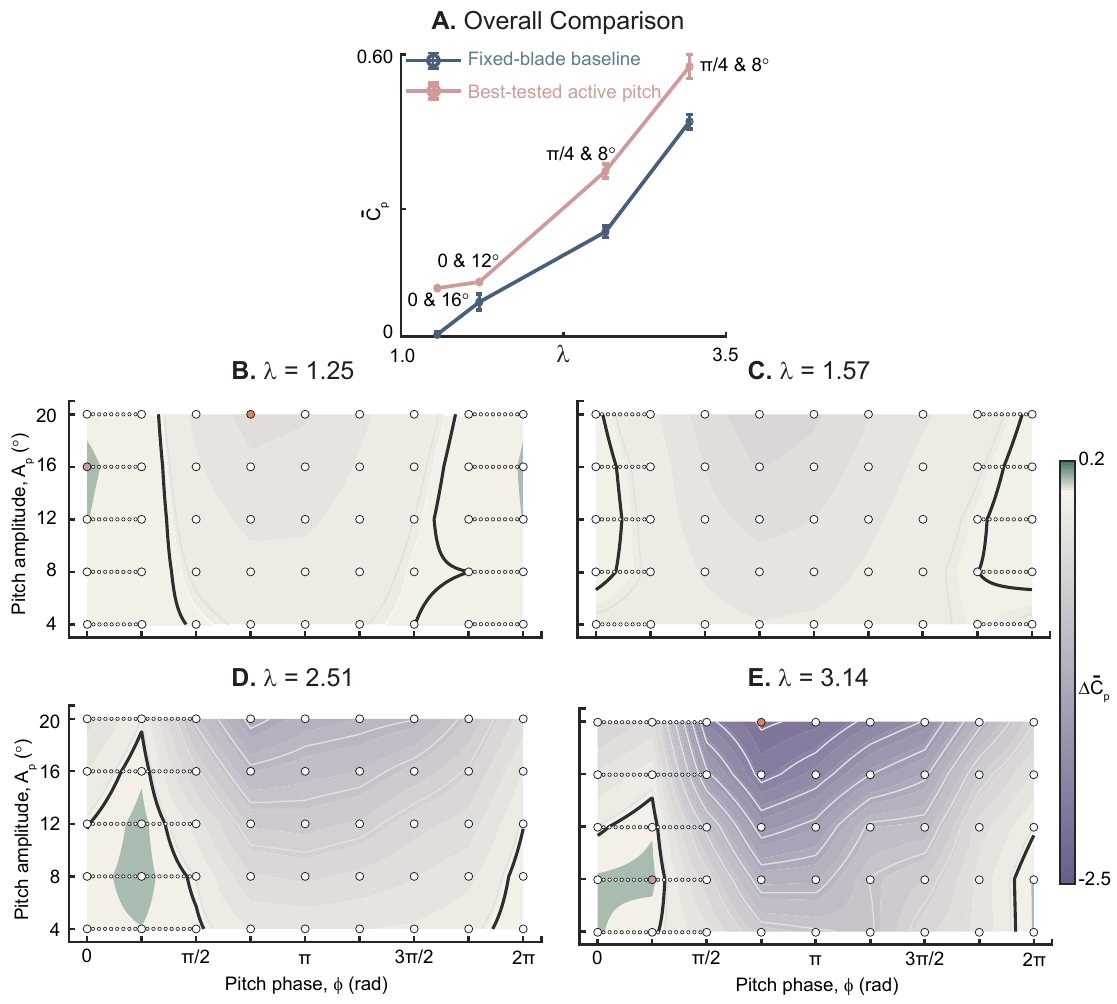}
        \end{subfigure}
    
	\caption{\textbf{Phase--amplitude dependence of active-pitch performance across tip-speed ratios.} \textbf{A.} Comparison of the cycle-averaged power coefficient, $\bar{C}_P$, between the fixed-blade baseline and the highest-performing tested active-pitch condition at $\lambda$ = 1.25, 1.57, 2.51, and 3.14. The corresponding pitch phase, $\phi$, and pitch amplitude, $A_p$, are indicated for each tip-speed ratio. \textbf{B-E.} Phase–amplitude distributions of the change in cycle-averaged power coefficient, $\Delta \bar{C}_P = \bar{C}_{P, \mathrm{active}} - \bar{C}_{P,\mathrm{fixed}}$, at \textbf{B.} $\lambda$ = 1.25, \textbf{C.} $\lambda$ = 1.57, \textbf{D.} $\lambda$ = 2.51, \textbf{E.} $\lambda$ = 3.14. Each contour map was constructed from the set of initial 40 experimentally tested phase–amplitude combinations (\textbf{large} white dots) at the corresponding tip-speed ratio. Together, the large and small white dots show all 120 tested phase--amplitude combinations. The dark-\textcolor{pink}{pink} and \textcolor{orange}{orange} filled circles identify the highest-performing and least favorable tested active-pitch conditions, respectively, selected at $\lambda$ = 1.25 and 3.14 for the phase-resolved comparison in Fig.~\ref{FIG:5}. The black contour denotes $\Delta \bar{C}_P = 0$, corresponding to active-pitch conditions that produce the same cycle-averaged power coefficient as the fixed-blade baseline.}
	\label{FIG:3}
\end{figure}

To quantify how the prescribed pitching kinematics affect turbine performance, the pitch phase, $\phi$, and amplitude, $A_p$, were varied systematically at each $\lambda$. Figure~\ref{FIG:3} summarizes the resulting change in cycle-averaged power coefficient relative to the corresponding fixed-blade baseline, $\Delta \bar{C}_P = \bar{C}_{P, \mathrm{active}} - \bar{C}_{P,\mathrm{fixed}}$, such that positive values represent an improvement in net power extraction produced by active pitching, whereas negative values indicate a reduction relative to the fixed-blade condition.

The phase--amplitude maps in Fig.~\ref{FIG:3}B--E demonstrate that the effect of active pitching depends strongly on both the timing and magnitude of the prescribed motion. Favorable and unfavorable regions coexist within the investigated parameter space at all four tip-speed ratios, indicating that active pitching does not inherently improve turbine performance~\cite{erickson2011performance}. The black $\Delta \bar{C}_P=0$ contours separate the combinations that outperform the fixed-blade baseline from those that reduce the cycle-averaged power coefficient. 

The highest-performing tested phase--amplitude combination at each tip-speed ratio is summarized in Fig.~\ref{FIG:3}A. At $\lambda=1.25$, the highest cycle-averaged performance was obtained with $\phi=0$ and a pitch amplitude of $A_p=16^\circ$, increasing $\bar{C}_P$ from 0.0047 for the fixed-blade baseline to 0.1037, corresponding to an improvement of approximately $2106\%$. At $\lambda=1.57$, the highest power was again obtained at $\phi=0$, with a reduced amplitude of $A_p=12^\circ$, increasing $\bar{C}_P$ from 0.0754 to 0.1173, or approximately $55.6\%$. At the two higher tip-speed ratios, the highest power was obtained at $\phi=\pi/4$ with $A_p=8^\circ$ at both $\lambda=2.51$ and $3.14$. The corresponding cycle-averaged power coefficients increased from 0.2289 to 0.3629 at $\lambda=2.51$ and from 0.4698 to 0.5948 at $\lambda=3.14$, representing improvements of approximately $58.5\%$ and $26.6\%$, respectively. The absolute values, including those near the Betz limit at $\lambda = 3.14$, reflect the confinement discussed in Section~2.1 and are interpreted only relative to the fixed-blade baseline. The exceptionally large relative improvement at $\lambda=1.25$ results in part from the near-zero fixed-blade baseline, although the absolute increase in $\bar{C}_P$ remains substantial. Li et al.~\cite{li2018optimization} also reported improved cycle-averaged performance with optimized active pitching for $\lambda<3.5$. These changes show that the amplitude and phase associated with the highest measured power depend on TSR. Similar TSR-dependent variations in optimal pitch amplitude and pitching schedule have been reported in previous variable-pitch VAWT studies~\cite{jain2016performance,le2024optimal}.

The extent of the favorable region indicates the range of pitch schedules that retain a performance benefit. At the lower tip-speed ratios, the favorable region extends over a relatively broad phase interval, whereas at higher $\lambda$ the positive-performance region becomes progressively narrower and more localized. In addition, larger pitch amplitudes increasingly fall within regions of negative $\Delta \bar{C}_P$ as $\lambda$ increases, indicating that excessive pitching becomes more detrimental at higher tip-speed ratios. Because the pitch phase is periodic, the favorable regions adjacent to $\phi=0$ and $2\pi$ represent a continuous phase window rather than two independent regions. 

The strong phase dependence observed in Fig.~\ref{FIG:3} further shows that pitch amplitude alone does not determine whether active pitching is beneficial. For a given $A_p$, varying $\phi$ can shift the turbine from enhanced power extraction to significant performance degradation. At a fixed TSR, this changes performance without increasing the prescribed pitch excursion or angular-rate magnitude. The energy-map analysis (Section~3.3) quantifies the associated net pitching-work exchange, while phase-resolved force and power measurements (Section~3.4) locate the gains and losses within the orbit.


\subsection{Turbine performance and pitching-energy exchange}

\label{sec:energy_map_results}

\begin{figure}
    \centering
    \includegraphics[width=0.9\textwidth]{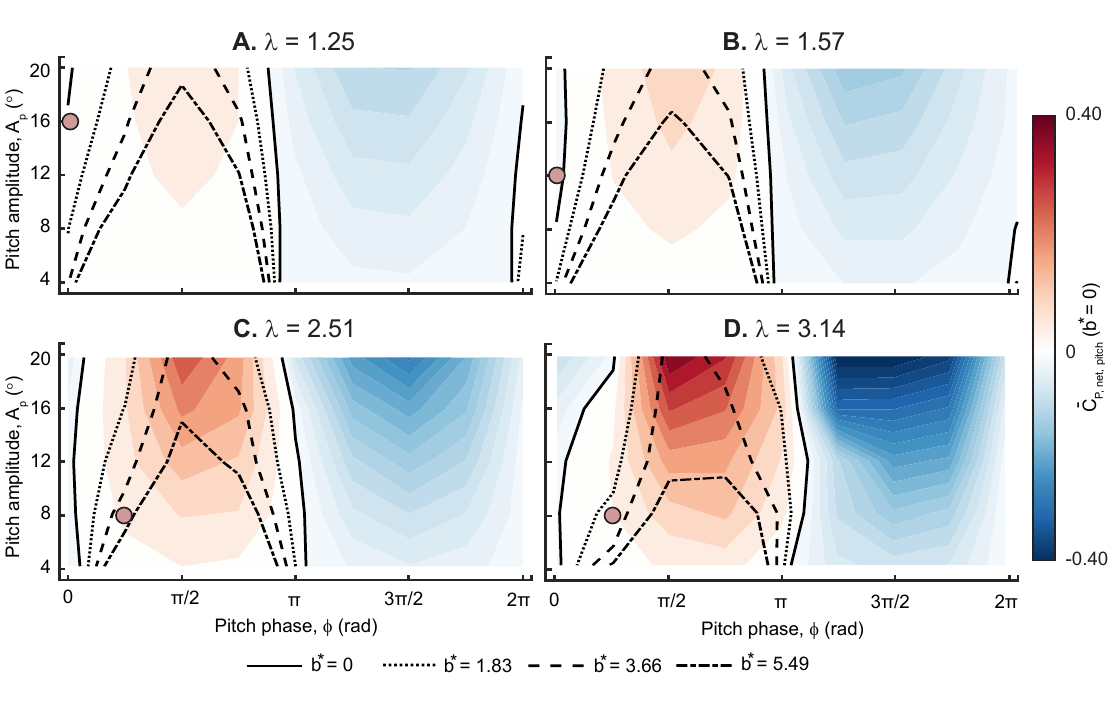}
    \caption{\textbf{Force-based energy maps of the pitching motion over the investigated tip-speed-ratio range.} \textbf{A--D.} Cycle-averaged net pitching power coefficient, $\bar{C}_{P,\mathrm{net,pitch}}$, at \textbf{A.} $\lambda = 1.25$, \textbf{B.} $\lambda = 1.57$, \textbf{C.} $\lambda = 2.51$, and \textbf{D.} $\lambda = 3.14$. The background contours show $\bar{C}_{P,\mathrm{net,pitch}}$ without damping ($b^{*} = 0$). Positive values (\textcolor{red}{red}) indicate net power transferred from the flow to the pitching degree of freedom, whereas negative values (\textcolor{blue}{blue}) indicate net energy transfer from the pitching motion to the flow. The black lines denote the $\bar{C}_{P,\mathrm{net,pitch}} = 0$ contours for $b^{*} = 0$ (solid), $1.83$ (dotted), $3.66$ (dashed), and $5.49$ (dash-dot), separating pitching conditions with positive and negative net pitching-power balances. Each contour is the locus of motions at which the hydrodynamic pitching work balances the corresponding damping loss, and hence of candidate steady states for a passive blade with that damping. Filled circles mark the highest-performing tested motion at each tip-speed ratio, as listed in Fig.~3A. For consistency across each tip-speed ratio, only the initial 40 phase--amplitude combinations were used to construct the maps, as these cases shared an identical parameter grid across all four operating conditions. The additional 80 refined cases were excluded because their tested phase ranges differed among tip-speed ratios.}
    \label{FIG:4}
\end{figure}

The pitching-power coefficients presented in this section use the blade planform area ($cs$) as the reference area; their corresponding values on the turbine swept-area basis are obtained by multiplying by $c/D$. The pitching-energy maps show, for each motion in Section~3.2, whether the flow does net work on the pitch axis and how much damping that work can cover. Figure~\ref{FIG:4} presents the resulting cycle-averaged net pitching power coefficient, $\bar{C}_{P,\mathrm{net,pitch}}$, over the same phase--amplitude space as Fig.~\ref{FIG:3}.

Without damping ($b^{*}=0$), the four maps exhibit a similar topology. $\bar{C}_{P,\mathrm{net,pitch}}$ is positive over approximately $0<\phi<\pi$ and negative over $\pi<\phi<2\pi$, with extrema near $\phi\approx\pi/2$ and $3\pi/2$. This pattern follows from the sinusoidal pitch law: because $\dot{\alpha}_p \propto \cos(\theta+\phi)$, only the first harmonic of $M_z(\theta)$ contributes to the cycle-averaged pitching work. To first order, this harmonic is set by the orbital incidence variation, so the work varies sinusoidally with $\phi$ at fixed amplitude. The extrema near $\phi \approx \pi/2$ and $3\pi/2$ indicate that this harmonic varies approximately as $\sin\theta$, consistent with the fixed-blade incidence $\alpha_{\mathrm{eff},0}(\theta)$ in Eq.~(1). The magnitude increases with both $A_p$ and $\lambda$, reaching approximately $0.40$ ($\phi \approx \pi/2$) and $-0.36$ ($\phi \approx 3\pi/2$) at $\lambda = 3.14$ and $A_p = 20^\circ$. Introducing virtual damping progressively contracts the region in which net hydrodynamic pitching work exceeds the assumed dissipation. Because the damper dissipation scales with $\dot{\alpha}_p^{2} \propto A_p^{2}\omega^{2}$, the region becomes bounded by a maximum pitch amplitude, which decreases with $\lambda$ for $b^{*} = 5.49$, from approximately $18.5^\circ$ at $\lambda = 1.25$ to $10.5^\circ$ at $\lambda = 3.14$. Each damped contour is the locus of candidate operating points for a passive blade with that damping (Section~2.5). At fixed phase, these points are stable in amplitude: below the contour the net energy input is positive and the amplitude grows, whereas above it the net input is negative and the amplitude decays, as in energy-map analyses of flow-induced pitching oscillations~\cite{zhu2020nonlinear,menon2019flow}. A complete stability assessment additionally requires the phase response, which depends on the moment component in phase with the pitch angle.

At $\lambda = 1.25$ and $1.57$, the highest-performing tested combinations ($\phi = 0$ with $A_p = 16^\circ$ and $12^\circ$) lie close to the zero-power contour for $b^{*} = 0$ and outside the $b^{*} = 1.83$ contour, so the cycle-averaged pitching work is nearly zero and smaller than the lowest assumed damping loss. At $\lambda = 2.51$ and $3.14$, the highest-performing tested condition ($\phi = \pi/4$, $A_p = 8^\circ$) lies within the positive region. It falls between the $b^{*} = 3.66$ and $5.49$ contours at $\lambda = 2.51$ and between the $b^{*} = 1.83$ and $3.66$ contours at $\lambda = 3.14$, so the damping at which the motion is energy-balanced lies within these ranges. At the two higher TSRs, the highest-performing motion can therefore cover a damping loss. At the two lower TSRs it cannot, but motions at slightly later phase lie inside the damped contours while still gaining power relative to the fixed blade (Figs.~3B,C and~4A,B), at a smaller gain.

However, the least favorable tested condition ($\phi = 3\pi/4$, $A_p = 20^\circ$) also lies within the positive region, despite producing the largest performance reduction at $\lambda = 3.14$ ($\Delta\bar{C}_P \approx -2.5$, Fig.~\ref{FIG:3}E). Because the normalizations in Eqs. \eqref{eq:net power} and \eqref{eq:bstar_Cpnetpitch} differ by the factor $c/D = 0.2$, the pitching motion accounts for at most approximately $0.08$ of $\bar{C}_P$, an order of magnitude smaller than the associated orbital power loss.

The maps exclude roughly half of the phase range, where the motion requires net power input, and give the damping at which each remaining motion is energy-balanced. They do not rank motions by power: the motion with the largest power loss at $\lambda = 3.14$ is also sustained by the flow. A passive design would additionally need the pitch inertia and stiffness that place the response at the intended phase. The following phase-resolved force and flow-field analyses examine how the prescribed motions produce these performance differences.

\subsection{Phase-resolved origins of power gains and losses}

\begin{figure}
	\centering
        \begin{subfigure}{1.0\textwidth}
            \centering
	       \includegraphics[width=0.5\linewidth]{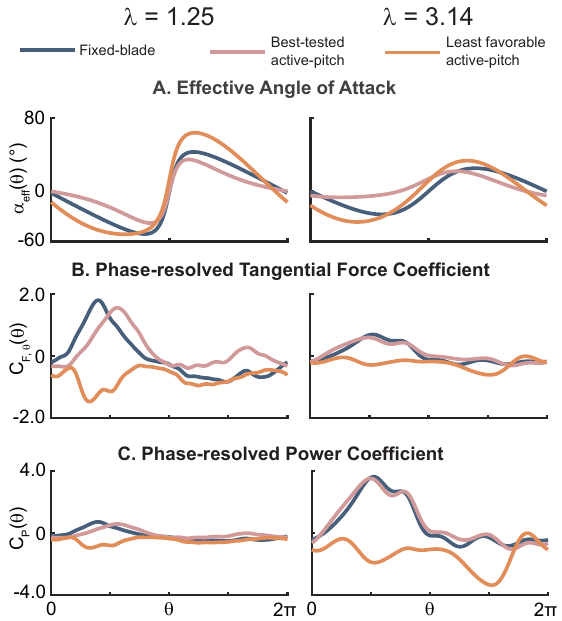}
        \end{subfigure}
    
	\caption{\textbf{Phase-resolved evolution of blade incidence, tangential loading, and power production for representative fixed-blade and active-pitch conditions.} The left and right columns correspond to $\lambda$ = 1.25 and 3.14, respectively. \textbf{A.} Effective angle-of-attack profiles, $\alpha_{\mathrm{eff}}(\theta)$, over one turbine revolution. \textbf{B.} Corresponding phase-resolved tangential force coefficients, $C_{F, \theta}(\theta)$. \textbf{C.} Corresponding phase-resolved power coefficients, $C_{P}(\theta)$.  \protect}
	\label{FIG:5}
\end{figure}

Changes in the timing and duration of positive and negative loading alter the balance of orbital work over the turbine cycle. Following the performance--energy comparison in Section~3.3, Fig.~\ref{FIG:5} compares the phase-resolved blade response for the fixed-blade baseline and the highest-performing and least favorable tested active-pitch conditions. The lowest and highest investigated tip-speed ratios, $\lambda=1.25$ and $3.14$, are selected as representative operating conditions to highlight the evolution of the active-pitch response across the tested range. At $\lambda=1.25$, the high-performing condition corresponds to $A_p=16^\circ$ and $\phi=0$, whereas the least favorable condition occurs at $A_p=20^\circ$ and $\phi=3\pi/4$. At $\lambda=3.14$, the corresponding high-performing condition is $A_p=8^\circ$ and $\phi=\pi/4$, while the least favorable condition again occurs at $A_p=20^\circ$ and $\phi=3\pi/4$, as identified in the contour maps in Fig.~\ref{FIG:3}. The comparison is organized through the effective angle of attack, tangential force coefficient, and instantaneous power coefficient, providing a direct link between the prescribed blade incidence, the resulting hydrodynamic loading, and power generation.

At $\lambda=1.25$, the high-performing active-pitch condition reduces the large incidence excursion relative to the fixed-blade baseline over both halves of the turbine cycle (Fig.~\ref{FIG:5}A). In contrast, the least favorable condition further increases the magnitude of $\alpha_{\mathrm{eff}}(\theta)$, particularly around the extrema of the fixed-blade trajectory. A similar distinction is also observed at $\lambda=3.14$, although the overall angle-of-attack variation is smaller owing to the higher blade orbital velocity. Chen et al.~\cite{chen2018numerical} showed that appropriately prescribed pitching can limit excessive angle-of-attack excursions and delay flow separation, but overly aggressive pitching can reduce aerodynamic performance. Sagharichi et al.~\cite{sagharichi2016variable} further showed that sinusoidal pitching systematically reduced the angle-of-attack range and delayed the occurrence of dynamic stall, but that excessive pitch amplitudes became detrimental to power production.

At $\lambda=1.25$, the high-performing motion shifts the dominant positive peak of $C_{F,\theta}(\theta)$ to a later azimuthal position relative to the fixed blade (Fig.~\ref{FIG:5}B). Although the peak tangential force is lower, favorable loading is redistributed toward later phases of the cycle, while the subsequent negative-loading excursion is reduced. Sagharichi et al.~\cite{sagharichi2016variable} reported a similar shift in peak torque with increasing pitch amplitude, linking the delay to reduced effective incidence and postponed stall development. Their results also showed that the azimuthal interval associated with high turbine torque ($F_{\theta}(\theta)R$) production can be extended through variable pitching. Similarly, Zhao et al.~\cite{zhao2018variable} demonstrated that variable pitching can improve VAWT performance by widening the azimuthal region of favorable torque production rather than simply increasing the maximum instantaneous turbine torque. At $\lambda=1.25$, the least favorable pitch schedule illustrates the opposite behavior: the large angle-of-attack excursion is accompanied by strongly negative $C_{F,\theta}$ over a considerable fraction of the cycle, substantially reducing the useful tangential loading. More generally, time-resolved measurements on fixed-blade VAWTs have shown that a large fraction of aerodynamic work at low TSR is produced during the shear-layer growth and dynamic-stall-vortex formation stages, emphasizing that the timing of stall development is as important as its occurrence~\cite{le2023time}.

At $\lambda=3.14$, the high-performing motion largely preserves the dominant positive-loading region while making smaller adjustments to its timing and subsequent negative loading. By contrast, the least favorable condition ($A_p=20^\circ$, $\phi = 3\pi/4$) suppresses much of the positive tangential loading and introduces unfavorable loading over the later portion of the cycle. The phase--amplitude maps show the same penalty (Fig.~\ref{FIG:3}). Excessive sinusoidal pitching can generate resistive torque, particularly during the downstream portion of the trajectory, as reported in numerical studies~\cite{chen2018numerical,chen2019improvement}. Chen et al.~\cite{chen2019improvement} further showed that reducing downstream pitching could recover power that would otherwise be lost to this resistive torque.

The cycle-averaged performance depends on the balance of positive and negative power over the full revolution, as illustrated by the histories in Fig.~\ref{FIG:5}C. At $\lambda=1.25$, the high-performing condition redistributes the favorable power-producing interval and moderates unfavorable contributions over the remainder of the cycle, without a substantially larger instantaneous power peak. At $\lambda=3.14$, the principal positive-power region remains similar between the fixed and high-performing cases, and the improvement again arises from comparatively modest but sustained changes throughout the cycle. The least favorable schedules generate extended negative-power intervals that reduce the cycle-integrated output. This distinction between increasing the peak response and extending the useful power-producing interval has also been emphasized in previous variable-pitch studies~\cite{zhao2018variable}.

The phase--amplitude combination changes $\alpha_{\mathrm{eff}}(\theta)$ and the timing and duration of favorable $C_{F,\theta}(\theta)$, redistributing $C_P(\theta)$ over the turbine cycle. This explains why the favorable pitching-energy balance discussed in Section~3.3 is insufficient to prevent performance losses when the orbital contribution is strongly reduced. The underlying flow structures responsible for these loading changes are examined subsequently through the phase-resolved flow-field measurements in Sections~3.5 and 3.6.

\subsection{Vortex-mediated power enhancement at the lowest TSR}

\begin{figure}
	\centering
        \begin{subfigure}{1.0\textwidth}
            \centering
	       \includegraphics[width=\linewidth]{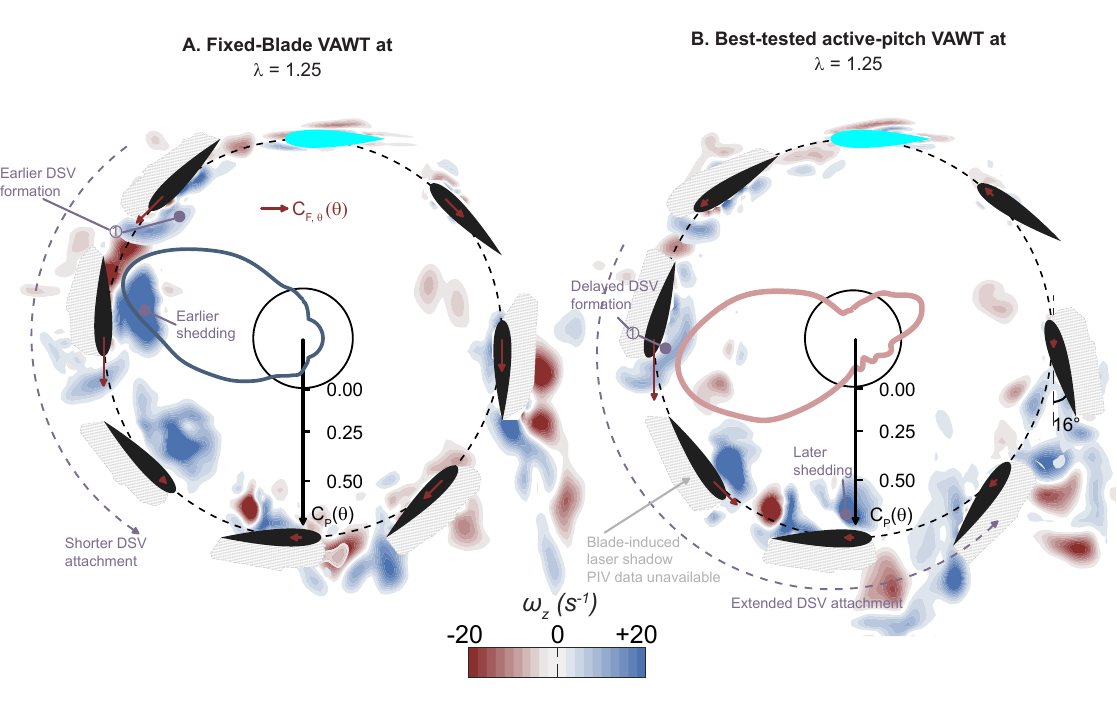}
        \end{subfigure}
    
	\caption{\textbf{Phase-resolved vortex evolution, blade loading, and power production for the fixed-blade and highest-performing tested active-pitch conditions at $\lambda$ = 1.25.} \textbf{A.} Fixed-blade baseline showing the phase-resolved spanwise vorticity field, $\omega_z$, together with the corresponding tangential force coefficient, $C_{F,\theta}(\theta)$, and instantaneous power coefficient, $C_{P}(\theta)$, over one turbine revolution. \textbf{B.} Corresponding active-pitch condition with $A_p = 16^\circ$ and $\phi = 0$. The phase-resolved PIV fields illustrate that active pitching delays dynamic-stall-vortex (DSV) formation, extends its attachment over a larger portion of the blade trajectory, and delays subsequent vortex shedding relative to the fixed-blade baseline. These changes coincide with a redistribution of the tangential loading and an extended favorable power-producing region, resulting in enhanced cycle-averaged energy extraction. Purple markers and annotations identify representative stages of DSV formation, attachment, and shedding. PIV data are unavailable over the indicated portion of the active-pitch trajectory because of blade-induced laser-sheet shadowing.  \protect}
	\label{FIG:6}
\end{figure}

Changes in dynamic-stall-vortex (DSV) development provide a flow-based explanation for the low-TSR loading redistribution identified in Section~3.4. Figure~\ref{FIG:6} compares DSV formation, attachment, and shedding between the fixed-blade and high-performing active-pitch conditions at $\lambda=1.25$.

For the fixed-blade case in Fig.~\ref{FIG:6}A, the large effective-angle-of-attack excursion at $\lambda=1.25$ is accompanied by relatively early DSV formation during the upstream portion of the trajectory. The growth of this coherent vortex coincides with the rapid increase and early peak in the positive tangential force observed in Fig.~\ref{FIG:5}B. However, the DSV remains associated with the blade over a comparatively short azimuthal interval before shedding earlier in the cycle. Following this shedding event, the favorable tangential loading decreases rapidly and is subsequently replaced by an extended region of negative $C_{F,\theta}$. Earlier time-resolved measurements likewise linked DSV development and shedding to the timing and magnitude of VAWT blade loading at low tip-speed ratios~\cite{le2022dynamic,le2023time}.

Active pitching delays DSV development and extends the blade--vortex association (Fig.~\ref{FIG:6}B). With $A_p=16^\circ$ and $\phi=0$, the prescribed pitch motion modifies the blade-incidence history during the upstream passage and delays DSV formation relative to the fixed-blade baseline. Once formed, the vortex remains associated with the blade over a larger portion of the trajectory and sheds at a later azimuthal position. This delayed and extended DSV evolution coincides with the later occurrence of the dominant positive $C_{F,\theta}$ response observed in Fig.~\ref{FIG:5}B and maintains favorable tangential loading farther into the turbine cycle. Similar benefits associated with controlling the timing of dynamic-stall development through blade pitching have been reported in previous variable-pitch VAWT studies~\cite{sagharichi2016variable,le2024optimal}.

The delayed DSV shedding coincides with a later reduction in favorable tangential loading and a smaller negative-power contribution following the primary loading event. Consequently, the favorable power-producing portion of the cycle is extended and redistributed toward later azimuthal positions. With nearly zero net pitching work (Section~3.3), the gain is attributed primarily to this redistribution of orbital loading.

\subsection{Flow-mediated performance enhancement at the highest TSR}

\begin{figure}
	\centering
        \begin{subfigure}{1.0\textwidth}
            \centering
	       \includegraphics[width=\linewidth]{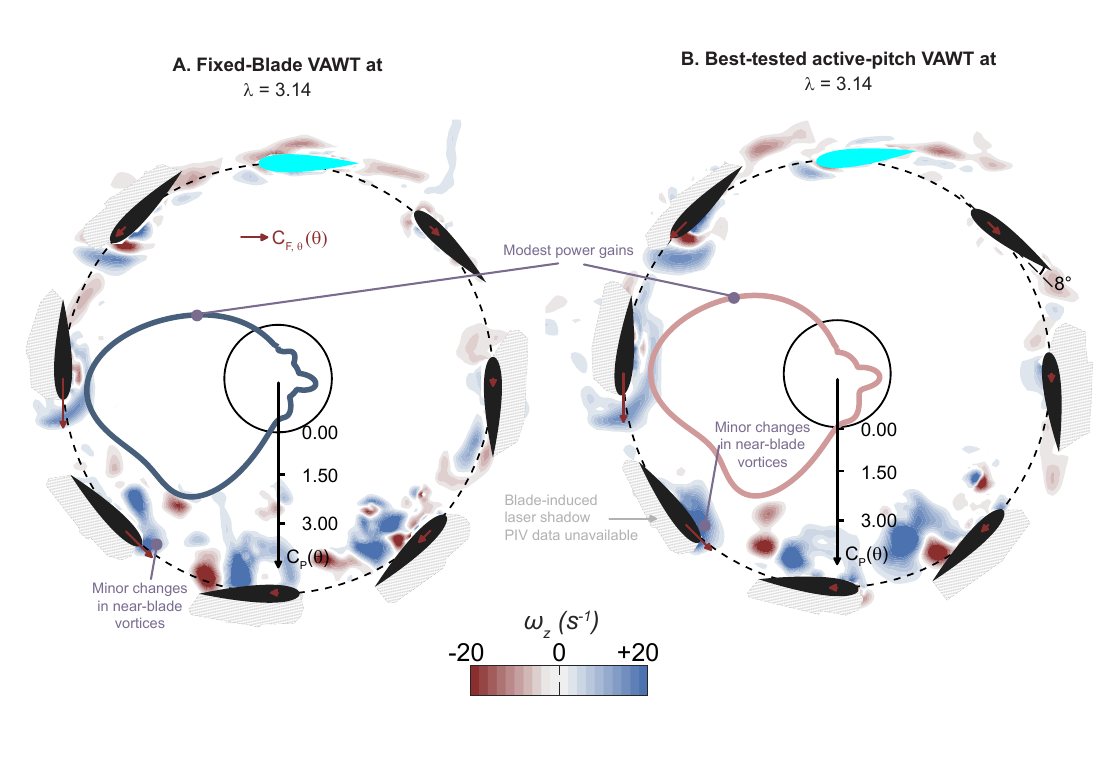}
        \end{subfigure}
    
	\caption{\textbf{Phase-resolved vortex evolution and power production for the fixed-blade and highest-performing tested active-pitch conditions at $\lambda$ = 3.14.} \textbf{A.} Fixed-blade baseline showing the phase-resolved spanwise vorticity field, $\omega_z$, together with the corresponding tangential force coefficient, $C_{F,\theta}(\theta)$, and instantaneous power coefficient, $C_{P}(\theta)$, over one turbine revolution. \textbf{B.} Corresponding active-pitch condition with $A_p = 8^\circ$ and $\phi = \pi/4$. Compared with the lowest-TSR condition, the active-pitch schedule produces only modest changes in the near-blade vortex structures and a correspondingly smaller redistribution of the phase-resolved loading and power production. Purple markers and annotations highlight representative regions of minor near-blade vortex modification and the associated power gain. PIV data are unavailable over the indicated portion of the trajectory because of blade-induced laser-sheet shadowing.  \protect}
	\label{FIG:7}
\end{figure}

At $\lambda=3.14$, the near-blade flow changes less under active pitching than at the lowest TSR. Figure~\ref{FIG:7} compares the fixed-blade baseline with the high-performing $A_p=8^\circ$, $\phi=\pi/4$ schedule, for which the positive-loading region remains largely intact (Section~3.4).

The near-blade vortex structures remain broadly similar between the fixed-blade and high-performing active-pitch cases (Fig.~\ref{FIG:7}). The smaller effective-angle-of-attack excursion at this tip-speed ratio limits the development of the pronounced dynamic-stall structures observed at $\lambda=1.25$, and the high-performing pitching motion produces only localized modifications to the near-blade vorticity field~\cite{le2024optimal}. The localized vorticity changes coincide with the phase-dependent loading adjustments in Fig.~\ref{FIG:5}, while the main positive-loading region remains largely intact. Together, these observations support a mechanism of localized flow adjustment that preserves the main power-producing event and moderates unfavorable contributions elsewhere in the cycle.

The comparison of Figs.~\ref{FIG:6} and \ref{FIG:7} shows that active pitching acts on different baseline flow regimes at the two TSRs. At $\lambda=1.25$, the large incidence excursion promotes strong dynamic-stall development, allowing active pitching to substantially alter DSV formation, attachment, and shedding and thereby extend favorable tangential loading. At $\lambda=3.14$, the baseline flow is less dominated by severe dynamic stall~\cite{le2022dynamic}, leaving less opportunity for active pitching to reorganize the vortex evolution. The corresponding highest-performing tested amplitudes are $16^\circ$ at $\lambda=1.25$ and $8^\circ$ at $\lambda=3.14$.

\section{Conclusions}
The present study demonstrates that the performance of a single-blade vertical-axis turbine model depends on the coupled effects of pitch amplitude, pitch phase, and tip-speed ratio. The experimental investigation of 480 prescribed pitching conditions shows that the same pitch amplitude can either improve or reduce power extraction depending on its timing relative to the blade orbital motion. Within the investigated parameter space, the pitch amplitude associated with the highest measured power decreased from $16^\circ$ at $\lambda = 1.25$ to $12^\circ$ at $\lambda = 1.57$, and to $8^\circ$ at $\lambda = 2.51$ and $3.14$, while the corresponding phase shifted from $0$ to $\pi/4$. 

The highest-performing tested motions are nearly neutral in net pitching work at the two lower TSRs and receive net energy from the flow through pitching at the two higher TSRs. With virtual damping, the latter retain a positive balance at $b^{*} = 1.83$, and at $b^{*} = 3.66$ for $\lambda = 2.51$. The motion with the largest power loss at $\lambda = 3.14$ also receives net energy from the flow, so a positive energy balance does not by itself indicate a power gain. At the two lower TSRs, motions at slightly later phase than the highest-performing one receive net energy from the flow while still gaining power, at a smaller gain.

Phase-resolved force and power measurements show that these performance gains are associated with changes in the timing and duration of favorable tangential loading. At the lowest tested tip-speed ratio, the high-performing pitching motion shifted the dominant positive-loading event toward later azimuthal positions and reduced subsequent unfavorable loading. The PIV observations indicate delayed dynamic-stall-vortex formation and shedding, with the vortex remaining close to the blade over a longer portion of the trajectory. At the highest tested tip-speed ratio, changes in the near-blade flow were localized and the dominant loading and power-production histories were largely preserved. 

Motions that both gain power and receive net energy from the flow are candidates for passive or assisted pitching; at the two higher TSRs they include the highest-performing tested motion. In principle, the damping and net stiffness of the pitch axis could be chosen so that a target motion becomes the steady passive response at a given operating condition: the maps provide the required damping, and the required stiffness and inertia follow from the moment component in phase with the pitch angle. Coupled fluid--structure experiments with prescribed pitch inertia, stiffness, and damping are needed to establish whether these motions arise and remain stable under passive operation.

\printcredits



\end{document}